\documentclass[twocolumn,showpacs,preprintnumbers,amsmath,amssymb]{revtex4}
\usepackage{graphicx}
\usepackage{dcolumn}
\usepackage{bm}
\usepackage[usenames]{color}

\newcommand{\psip}{\psi^{\prime}}
\newcommand{\chicJ}{\chi_{cJ}}
\newcommand{\chicz}{\chi_{c0}}

\newcommand{\chict}{\chi_{c2}}

\begin{document}

\title{\boldmath Study of $\chi_{cJ}$ radiative decays into a vector meson}
\author{M.~Ablikim$^{1}$, M.~N.~Achasov$^{5}$, D.~Alberto$^{38}$,
L.~An$^{9}$, Q.~An$^{36}$, Z.~H.~An$^{1}$, J.~Z.~Bai$^{1}$,
R.~Baldini$^{17}$, Y.~Ban$^{23}$, J.~Becker$^{2}$
% ...
\\
\vspace{0.2cm}
(BESIII Collaboration)\\
\vspace{0.2cm}
{\it
$^{1}$ Institute of High Energy Physics, Beijing 100049, P. R. China\\
$^{2}$ Bochum Ruhr-University, 44780 Bochum, Germany\\
% ...
$^{5}$ G.I. Budker Institute of Nuclear Physics SB RAS (BINP), Novosibirsk 630090, Russia\\
% ...
$^{9}$ Guangxi University, Naning 530004, P. R. China\\
% ...
$^{17}$ Laboratori Nazionali di Frascati - INFN, 00044 Frascati, Italy\\
% ...
$^{23}$ Peking University, Beijing 100871, P. R. China\\
% ...
$^{36}$ University of Science and Technology of China, Hefei 230026, P. R. China\\
% ...
$^{38}$ University of Turin and INFN, Turin, Italy\\
% ...
}} \vspace{0.6cm}

\date{\today}

\begin{abstract}
The decays $\chi_{cJ}\to\gamma V$ ($V=\phi,~\rho^0,~\omega$) are
studied with a sample of radiative $\psip\to\gamma\chi_{cJ}$ events
in a sample of $(1.06\pm0.04)\times 10^{8}~\psip$ events collected
with the BESIII detector. The branching fractions are determined to
be: ${\cal B}(\chi_{c1}\to \gamma\phi)=(25.8\pm 5.2\pm 2.3)\times
10^{-6}$, ${\cal B}(\chi_{c1}\to \gamma\rho^0)=(228\pm 13\pm
22)\times 10^{-6}$, and ${\cal B}(\chi_{c1}\to
\gamma\omega)=(69.7\pm 7.2\pm 6.6)\times 10^{-6}$. The decay
$\chi_{c1}\to \gamma\phi$ is observed for the first time.  Upper
limits at the 90\% confidence level on the branching fractions for
$\chi_{c0}$ and $\chict$ decays into these final states are
determined. In addition, the fractions of the transverse
polarization component of the vector meson in $\chi_{c1}\to \gamma
V$ decays are measured to be $0.29_{-0.12-0.09}^{+0.13+0.10}$ for
$\chi_{c1}\to \gamma\phi$, $0.158\pm 0.034^{+0.015}_{-0.014}$ for
$\chi_{c1}\to \gamma\rho^0$, and
$0.247_{-0.087-0.026}^{+0.090+0.044}$ for $\chi_{c1}\to
\gamma\omega$, respectively. The first errors are statistical and
the second ones are systematic.
\end{abstract}

\maketitle
\section{Introduction}
Doubly radiative decays of the type $\psi\to\gamma X\to \gamma\gamma
V$, where V is either a $\phi$, $\rho^0$, or $\omega$ meson, provide
information on the flavor content of the $C$-even resonance $X$ and
on the gluon hadronization dynamics in the
process~\cite{zoubs,zhaogd,zhaogd1}. The spin and charge dependent
couplings in radiative decays reveal detailed information which is
particularly useful in the search for glueball and hybrid
states~\cite{close2002}. For the case where $X = \chicJ$, the decay
of the $P$-wave $\chicJ$ to $\gamma V$ may provide an independent
window for understanding possible glueball dynamics and validating
theoretical techniques~\cite{close1996}.

Table~\ref{tab1} shows the theoretical predictions for $\chicJ$
radiative decays to a vector meson from perturbative quantum
chromodynamics (pQCD)~\cite{zhaogd}, nonrelativistic QCD
(NRQCD)~\cite{zhaogd1}, and NRQCD plus QED contributions
(NRQCD+QED)~\cite{zhaogd1}, and recent results from the CLEO
experiment~\cite{cleoc}. The experimental results for ${\cal
B}(\chi_{c1}\to \gamma \rho^0$, $\gamma\omega)$ are an order of
magnitude higher than the corresponding theoretical predictions.
However, by including non-perturbative QCD hadronic loop
contributions, a recent pQCD calculation~\cite{chen} obtains results
in agreement with the CLEO measurements.  Improved measurements of
$\chicJ$ radiative decays to vector mesons using the large BESIII
$\psip$ sample will provide tighter constraints on theoretical
calculations.

In this paper, we present measurements of radiative decays of the
$\chicJ$ to the light vector mesons. The measurements have improved
precision compared to CLEO's results, and $\chi_{c1} \rightarrow
\gamma \phi$ decay is observed for the first time. In addition, the
fraction of the transverse polarization component of the vector meson
in $\chi_{c1}\to \gamma V$ decay is studied, and the results indicate
that the longitudinal component for $\chi_{c1} \rightarrow \gamma V$
decay is dominant. This observation may help in the theoretical
understanding of $\chi_{c1} \rightarrow \gamma V$ decays.

\begin{table}[hbtp]
\caption{\label{tab1} Comparison of theoretical predictions on the
branching fractions for $\chicJ$ radiative decays to a vector
meson (in units of $10^{-6}$) and measurements from the CLEO
experiment. The upper limits are at the 90\% confidence level
(C.L.).}
\begin{ruledtabular}
\begin{tabular}{cccccc}
    Mode & CLEO~\cite{cleoc}   & pQCD~\cite{zhaogd}  & NRQCD~\cite{zhaogd1}
            & NRQCD\\
         &&&&+QED~\cite{zhaogd1}\\ \hline
 $\chi_{c0}\to\gamma\rho^{0}$ & $<9.6$               & 1.2 &3.2 &2.0 \\
 $\chi_{c1}\to\gamma\rho^{0}$ & $243\pm 19\pm 22$    & 14  &41  &42  \\
 $\chi_{c2}\to\gamma\rho^{0}$ & $<50$                & 4.4 &13  &38  \\
 \hline
 $\chi_{c0}\to\gamma\omega$ &  $<8.8$             & 0.13 &0.35&0.22 \\
 $\chi_{c1}\to\gamma\omega$ & $83\pm 15\pm 12$    & 1.6  &4.6 &4.7  \\
 $\chi_{c2}\to\gamma\omega$ &  $<7.0$             & 0.5  &1.5 &4.2  \\
 \hline
 $\chi_{c0}\to\gamma\phi$ &  $<6.4$               & 0.46&1.3&0.03   \\
 $\chi_{c1}\to\gamma\phi$ &  $<26$                & 3.6&11&11       \\
 $\chi_{c2}\to\gamma\phi$ &  $<13$                & 1.1&3.3&6.5     \\
\end{tabular}
\end{ruledtabular}
\end{table}

\section{BEPCII collider and BESIII detector}

BEPCII/BESIII~\cite{bes3} is a major upgrade of the BESII experiment
at the BEPC accelerator~\cite{bes2} for studies of hadron spectroscopy
and $\tau$-charm physics \cite{bes3phys}. The design peak luminosity
of the double-ring $e^+e^-$ collider, BEPCII, is $10^{33}$
cm$^{-2}$s$^{-1}$ at a beam current of 0.93~A.  The BESIII detector
with a geometrical acceptance of 93\% of 4$\pi$, consists of the
following main components: 1) a small-celled, helium-based main draft
chamber (MDC) with 43 layers.  The average single wire resolution is
135 $\mu$m, and the momentum resolution for 1~GeV/$c$ charged
particles in a 1 T magnetic field is 0.5\%; 2) an electromagnetic
calorimeter (EMC) made of 6240 CsI (Tl) crystals arranged in a
cylindrical shape (barrel) plus two end-caps.  For 1.0~GeV photons, the
energy resolution is 2.5\% in the barrel and 5\% in the end-caps, and
the position resolution is 6~mm in the barrel and 9~mm in the end-caps;
3) a Time-Of-Flight system (TOF) for particle identification composed
of a barrel part made of two layers with 88 pieces of 5~cm thick, 2.4~m
long plastic scintillators in each layer, and two end-caps with 96
fan-shaped, 5~cm thick, plastic scintillators in each end-cap.  The
time resolution is 80~ps in the barrel, and 110 ps in the end-caps,
corresponding to a $2\sigma$ K/$\pi$ separation for momenta up to
about 1.0~GeV/$c$; 4) a muon chamber system (MUC) made of 1000~m$^2$
of Resistive Plate Chambers (RPC) arranged in 9 layers in the barrel
and 8 layers in the end-caps and incorporated in the return iron of the
superconducting magnet.  The position resolution is about 2~cm.

The optimization of the event selection and the estimation of
physics backgrounds are performed through Monte Carlo simulations.
The GEANT4-based simulation software BOOST~\cite{boost} includes the
geometric and material description of the BESIII detectors, the
detector response and digitization models, as well as the tracking
of the detector running conditions and performance.  The production
of the $\psip$ resonance is simulated by the Monte Carlo event
generator KKMC~\cite{kkmc}, while the decays are generated by
EvtGen~\cite{besevtgen} for known decay modes with branching ratios
being set to the PDG~\cite{pdg10} world average values, and by
Lundcharm~\cite{lund} for the remaining unknown decays.  The
analysis is performed in the framework of the BESIII Offline
Software System~(BOSS)~\cite{boss} which takes care of the detector
calibration, event reconstruction and data storage.

\section{Data Analysis}

A data sample of $(1.06\pm 0.04)\times 10^{8}$ $\psip$ events
collected with the BESIII detector is used in this analysis, and an
independent sample of about 42.6~pb$^{-1}$ taken at
$\sqrt{s}=3.65$~GeV is utilized to determine the potential
background contribution from the continuum. In this paper, we focus
on the exclusive decays of $\psip\to \gamma_l \chi_{cJ}$,
$\chi_{cJ}\to \gamma_h V$, where $\gamma_l (\gamma_h)$ designates
the lower (higher) energy photon and V is either a $\phi$, $\rho^0$,
or $\omega$ meson. The $\phi$, $\rho^0$, and $\omega$ candidates are
reconstructed in the $K^+K^-$, $\pi^+\pi^-$, and $\pi^+\pi^-\pi^0$
decay modes, respectively.

Charged tracks are reconstructed in the MDC, and the number of
charged tracks is required to be two with net charge zero. For each
track, the polar angle must satisfy $|\cos\theta|< 0.93$, and it
must be within $\pm 10$~cm of the interaction point in the beam
direction and within $\pm 1$~cm of the beam line in the plane
perpendicular to the beam. Since the efficiency of particle
identification (PID) is lower for higher momentum ($>1$ GeV/$c$)
charged tracks, only the lower momentum charged track is required to
be identified as a $K$ or $\pi$.

Electromagnetic showers are reconstructed by clustering EMC crystal
energies. The energy deposited in nearby TOF counters is included to
improve the reconstruction efficiency and energy resolution. Showers
identified as photon candidates must satisfy fiducial and
shower-quality requirements.  The photon candidate showers
reconstructed from the barrel region $(|\cos \theta|<0.8)$ must have a
minimum energy of 25~MeV, while those in the end-caps $(0.86<|\cos
\theta|<0.92)$ must have at least 50~MeV. The showers in the angular
range between the barrel and end-cap are poorly reconstructed and
excluded from the analysis. To eliminate showers from charged
particles, a photon must be separated by at least $10^{\circ}$ from
any charged track. EMC cluster timing requirements are used to
suppress electronic noise and energy deposits unrelated to the event.

In order to choose the correct combination and improve the mass
resolution, a four-constraint kinematic fit (4C-fit) is done under the
assumption of energy-momentum conservation. Candidates with
$\chi^2\leq 100$ for this fit are retained.  If an event has more than
one candidate, the candidate with the smallest $\chi^2$ is kept.

For $\psip\to\gamma\gamma \omega$ ($\omega \rightarrow
\pi^+\pi^-\pi^0$) candidates, the gammas from the $\pi^0$ decay are
selected as those that give the minimum of
$$\sqrt{\left(\frac{M_{\gamma_1\gamma_2}-M_{\pi^0}}{\sigma_{\pi^0}}\right)^2
+\left(\frac{M_{\pi^+\pi^-\gamma_1\gamma_2}-M_\omega}{\sigma_{\omega}}\right)^2}~,
$$ where $M_{\gamma_1\gamma_2}$ is the invariant mass of the photon
pair, $M_{\pi^0}$ ($M_\omega$) is the nominal mass of $\pi^0$
($\omega$), $\sigma_{\pi^0}$ ($\sigma_{\omega}$) is the mass
resolution determined from MC simulation and is about 7~MeV/$c^2$ (
6~MeV/$c^2$). A $\pi^0$ mass constraint for the $\gamma\gamma\omega$
channel is included by doing a five-constraint kinematic fit (5C-fit),
and events with $\chi^2_{5C}\leq 100$ are kept as $\psip\to
\gamma_l\gamma_h\omega$ ($\omega \rightarrow \pi^+\pi^-\pi^0$)
candidates.

To suppress background from multi-photon hadronic decays of the
$\psip$, $|M_{\gamma_l\gamma_h}-M_{\eta}|\geq 25 {\rm ~MeV}/c^2$ for
$\gamma\gamma \phi$, $M_{\gamma_l\gamma_h}\geq 600 {\rm ~MeV}/c^2$ for
$\gamma\gamma \rho^0$, and $|M_{\gamma_l\gamma_h}-M_{\eta}|\geq 25
{\rm ~MeV/c^2}$ and $|M_{\gamma_l\gamma_h}-M_{\pi^0}|\geq 15 {\rm
~MeV}/c^2$ for $\gamma\gamma \omega$ are required. Here $M_{\pi^0}$
and $M_{\eta}$ are the nominal masses of $\pi^0$ and $\eta$,
respectively. The background from $\psip\to \gamma_h\eta^\prime$,
$\eta^\prime\to \gamma_l V$ ($V=\rho^0$, $\omega$) is suppressed by
requiring $|M_{\gamma_l V}-M_{\eta^\prime}|>15 {\rm ~MeV}/c^2$.

In $\chi_{cJ}\to \gamma\rho^0$, there are potential backgrounds from
QED $e^+e^- \rightarrow \gamma e^+e^-$ and $\mu^+\mu^-$ events where
the leptons are misidentified as pions. To reject electrons, the
ratio of the energy deposited in the EMC to the momentum measured in
the MDC ($E_{\rm{EMC}}/cp_{\rm{MDC}}$) of tracks must be less than
0.8. To reject muons, tracks are removed if the number of layers
with hits in the muon chamber is greater than three. The QCD
backgrounds remaining can be effectively eliminated by requiring the
opening angle between the two pions, $\cos\theta_{\pi^+\pi^-}$,
satisfy $\cos\theta_{\pi^+\pi^-} >-0.8$, and that between the two
photons, $\cos \theta_{\gamma_l\gamma_h}$, satisfy $-0.98< \cos
\theta_{\gamma_l\gamma_h} < 0.5$, in the laboratory frame.

\begin{figure}[htbp]
  \includegraphics[width=0.35\textwidth,height=2.2cm,angle=0]{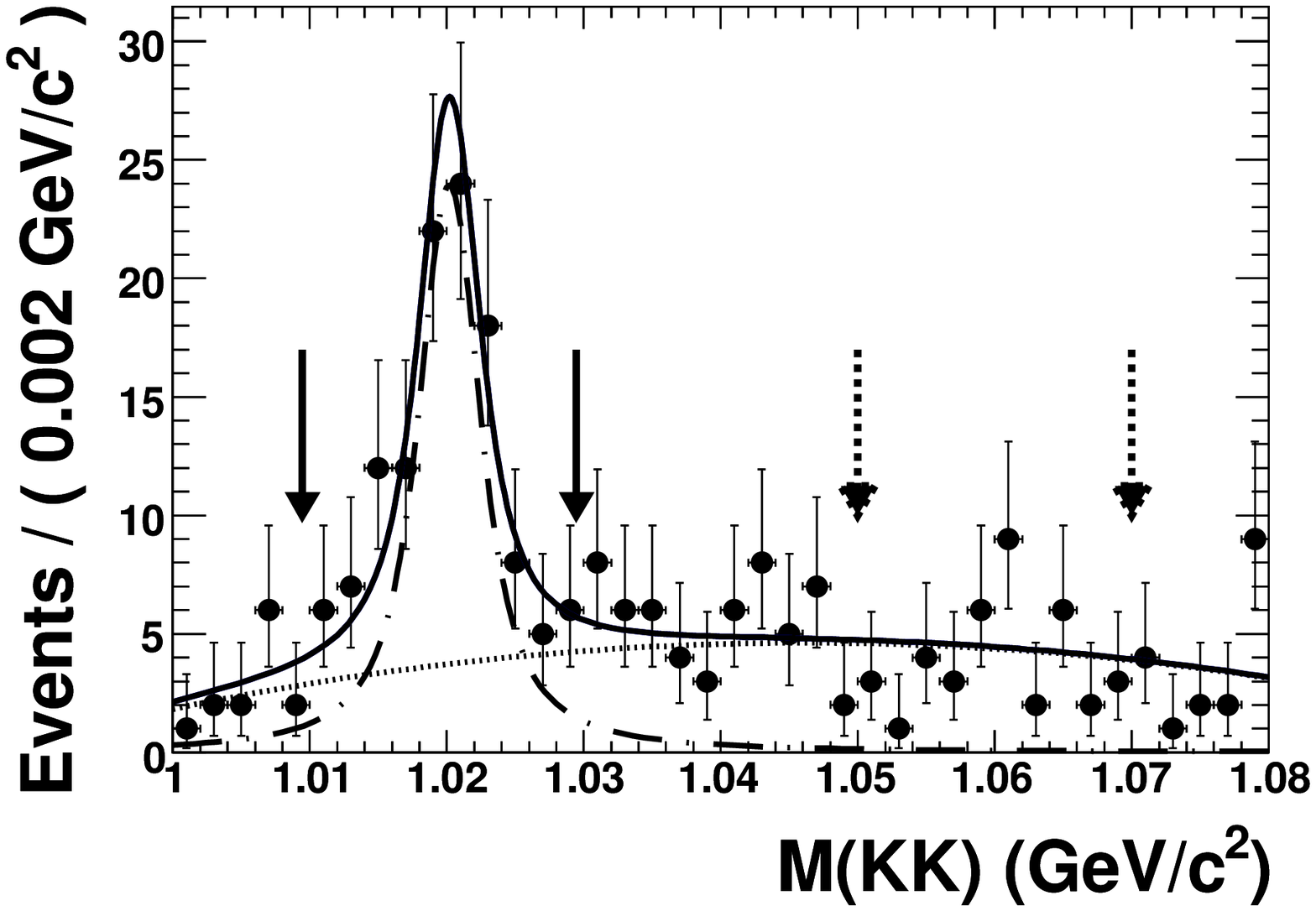}
             \put(-70,45){(a)}
   \par
   \includegraphics[width=0.35\textwidth,height=2.2cm,angle=0]{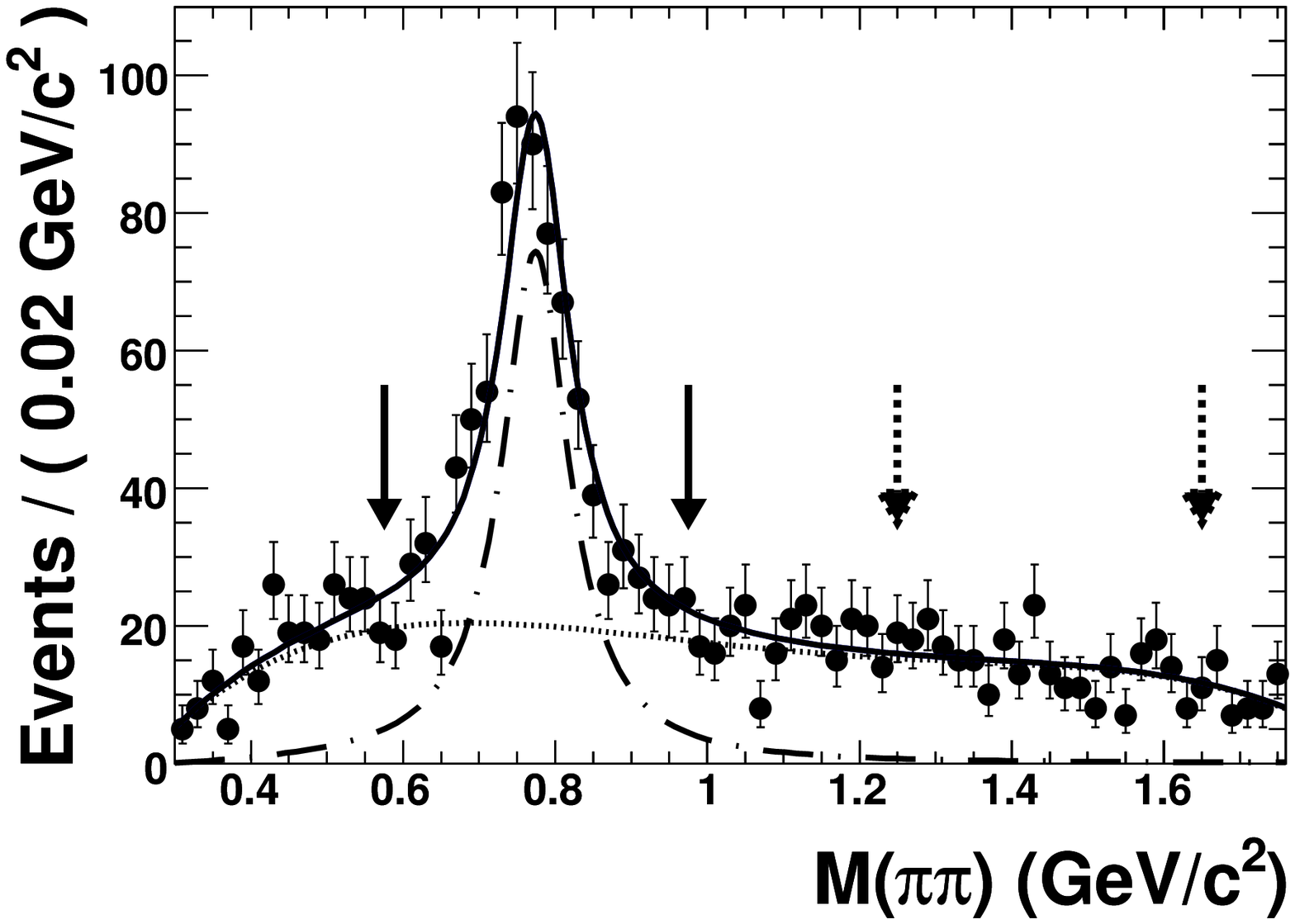}
              \put(-70,45){(b)}
   \par
   \includegraphics[width=0.35\textwidth,height=2.2cm,angle=0]{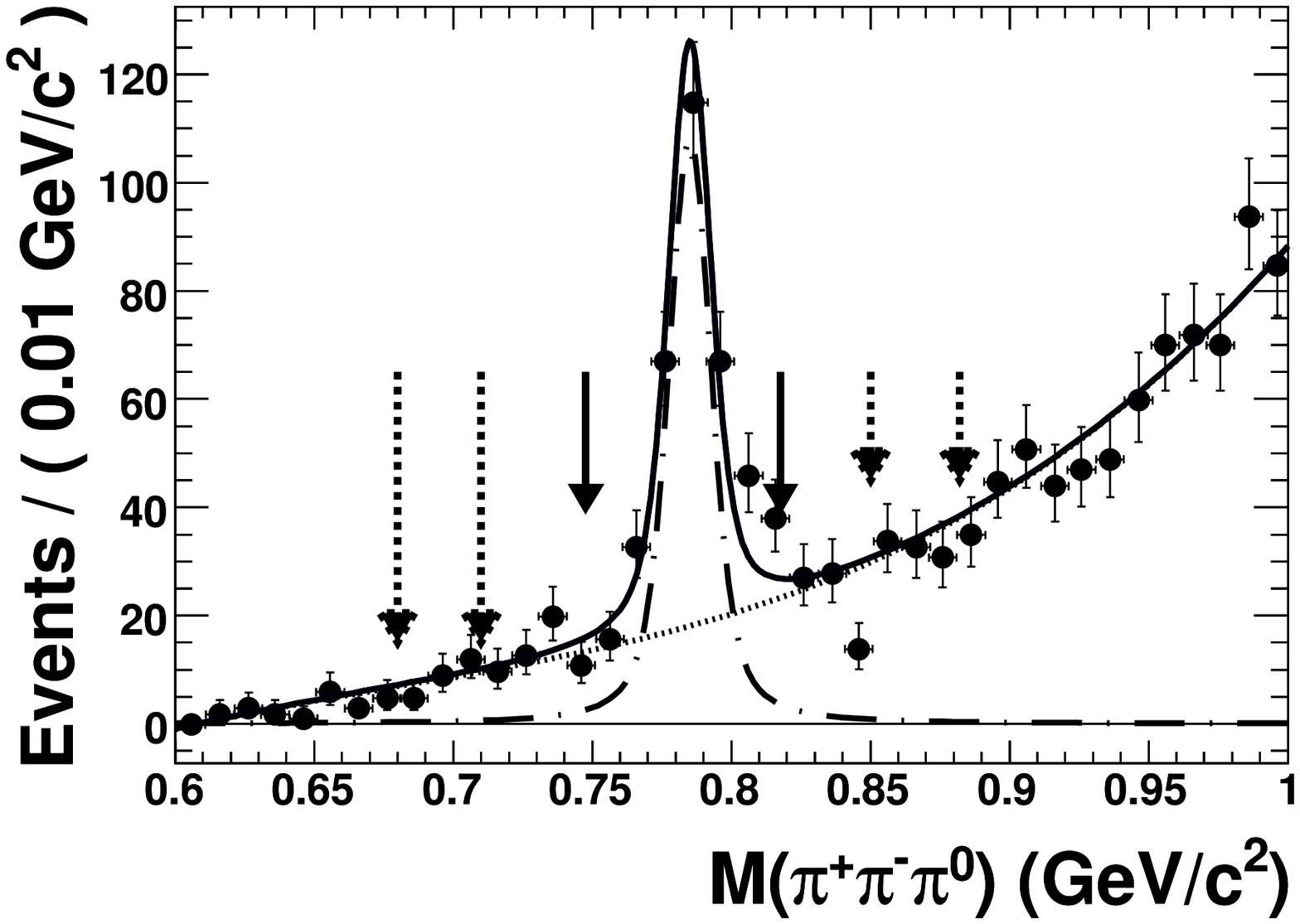}
              \put(-70,45){(c)}
              \par
\caption{\label{fig:mv}Invariant mass distributions of (a) $K^+K^-$,
(b) $\pi^+\pi^-$, and (c) $\pi^+\pi^-\pi^0$. Dots with error bars are
data; dashed lines are signal shapes; and dotted lines are the
polynomial background contributions. The signal regions and sideband
regions are indicated with the solid and dashed arrows respectively.}
\end{figure}

Figure~\ref{fig:mv} shows the $K^+K^-$, $\pi^+\pi^-$, and
$\pi^+\pi^-\pi^0$ invariant mass distributions for the candidate
events. The curves show the best fit to the mass spectra using a
$s$-dependent Breit-Wigner function for signal and a polynomial for
background. Events with $|M_{K^+K^-}-M_{\phi}|\le 0.01$~GeV/$c^2$,
$|M_{\pi^+\pi^-} - M_\rho|\le 0.2 {\rm GeV}/c^2$, and
$|M_{\pi^+\pi^-\pi^0} - M_\omega|\le 0.035 {\rm GeV}/c^2$ are taken
as $\phi$, $\rho^0$, and $\omega$ candidates, respectively. Here
$M_{\phi}$, $M_\rho$, and $M_\omega$ are the nominal masses of these
vector mesons. The sideband regions are defined as $1.05\le
M_{K^+K^-}\le 1.07 {\rm GeV}/c^2$, $1.25\le M_{\pi^+\pi^-}\le 1.65
{\rm GeV}/c^2$, and ($0.68\le M_{\pi^+\pi^-\pi^0}\le 0.71{\rm
GeV}/c^2$ and $0.85\le M_{\pi^+\pi^-\pi^0}\le 0.88{\rm GeV}/c^2$)
for the $\phi$, $\rho^0$, and $\omega$, respectively.

\begin{figure}[htbp]
   \centering
   \includegraphics[width=0.35\textwidth,height=6.6cm,angle=0]{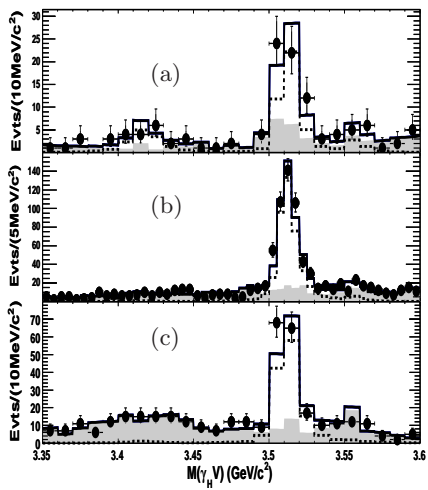}
             \put(-120,155){(a)}
             \put(-120,105){(b)}
             \put(-120,55){(c)}

\caption{\label{fig:mgv}Invariant mass distributions of (a)
$\gamma\phi$, (b) $\gamma\rho^0$, and (c) $\gamma\omega$. Dots with
error bars are data; histograms are the best fit; dashed histograms
are signal shapes; and the gray-shaded histograms are the sum of the
sideband background and the background polynomial.}
\end{figure}

After applying the above criteria, there are still several peaking
backgrounds in the $\chi_{cJ}$ mass regions from $\chi_{cJ}$ decays
into non-$\gamma V$ modes with the same final states. From MC
studies, the shapes of these backgrounds are found to be similar to
those of the vector meson sideband background events. The invariant
mass distributions of $\gamma_{h} V$, where $V=\phi$, $\rho^0$,
$\omega$, respectively, are shown in
Figs.~\ref{fig:mgv}(a)-\ref{fig:mgv}(c). There are clear $\chi_{c1}$
signals in all decay modes, while $\chicz$ and $\chict$ signals are
not evident. In order to extract the signal yields from the mass
spectra, we first obtain signal shapes for each $\chi_{cJ}\to \gamma
V$ mode (9 decay modes in total) using MC simulations. Each of the
distributions in Fig.~\ref{fig:mgv} is fitted with a background
shape composed of the vector meson mass sideband distribution plus a
2nd order polynomial function and three $\chi_{cJ}$ resonances as
the signal shapes. Parameters of the polynomial function and the
normalization for each of the $\chi_{cJ}$ resonances are allowed to
float in the fit. The fitted yields are summarized in
Table~\ref{tabtotal}.  $\chi_{c1} \rightarrow \gamma \rho$ and
$\gamma \omega$ are observed with a statistical significance larger
than $10 \sigma$, and the significance for $\chi_{c1} \rightarrow
\gamma \phi$ is $6.4 \sigma$. Here, the significance is determined
from $\sqrt{-2 \rm{log}({\cal L}_0/{\cal L}_{\rm{max}})}$, where
${\cal L}_{\rm{max}}$ is the maximum likelihood value, and ${\cal
L}_0$ is the likelihood for a fit with the signal contribution set
to zero. Branching fractions are calculated after considering the
signal efficiency, as listed in Table~\ref{tabtotal}, and the upper
limits at the 90\% C.L on the branching factions of $\chicz$ and
$\chict$ decays are estimated by a Bayesian method~\cite{up}. The
effects of both the statistical and systematic uncertainties to the
upper limits are taken into account. All results are listed in
Table~\ref{tabtotal}.

\section{Estimation of systematic uncertainties}

Table~\ref{tabsys} shows the summary of all sources of systematic
uncertainties. Many systematic uncertainties are determined using
clean, high statistics control samples that allow results from MC
simulation to be compared with those from data.
\par

\subsection{Photon efficiency\label{photon_effi_err}}
The photon detection efficiency and its uncertainty are studied by
three different methods.
\par
The missing photon method uses a sample of $\psi^\prime\to\pi^+\pi^-
J/\psi$, $J/\psi\to \rho^0\pi^0$ events. Using events with four
charged tracks, identified as pions, plus a good photon, the missing
momentum is determined and used to predict the direction and energy of
the missing photon.  To remove background, the invariant mass of the
good photon and the missing momentum must be consistent with that of
the $\pi^0$, and the invariant mass of the charged tracks from the
$J/\psi$ decay must be consistent with that of the $\rho$.  The photon
detection efficiency is then the fraction of actual photons matched in
direction to the predicted photon. On average, the efficiency
difference between data and Monte Carlo simulation is less than 1\%.
\par
The missing $\pi^0$ method uses a sample of
$\psi^\prime\to\pi^0\pi^0 J/\psi$, $J/\psi\to l^+l^-$ events, in a
similar way to the first method. Events with two charged tracks and
at least two photons are required. The invariant mass of the charged
tracks must be consistent with the mass of the $J/\psi$, and the sum
of the momenta of the two photons must be greater than 300 MeV/$c$.
Since the two pions are anti-correlated, by requiring a $\pi^0$ with
momentum larger than 300 MeV/$c$ and using energy momentum
conservation, the energy and direction of the remaining soft $\pi^0$
can be predicted.  To ensure a clean sample, the invariant mass of
the two photons of the selected pion and the mass recoiling against
the $J/\psi$ and the selected $\pi^0$ must be consistent with that
of the $\pi^0$.  The number of reconstructed $\pi^0$s, which match
in direction and have two photon invariant mass consistent with the
$\pi^0$ mass, yields the $\pi^0$ detection efficiency. The
difference in efficiency between the data and Monte Carlo is less
than $(1.5\pm 0.5)\%$; hence the uncertainty of photon detection is
less than 1\%.
\par
The third method, the $\pi^0$ decay angle method~\cite{bes2photon},
utilizes a $J/\psi\to\rho^0\pi^0$ sample. Since the $\pi^0$ is from
a two-body decay, the momentum of the $\pi^0$ is known, and the
energy of the lower energy daughter photon is given by $2E_{low} =
\sqrt{P^2_{\pi^0}+ M^2_{\pi^0}}-P_{\pi^0} \cos\theta$, where
$\theta$ is the angle of the photon in the $\pi^0$ rest frame with
respect to the direction of the $\pi^0$ in the $J/\psi$ rest frame.
For spin-zero particles like the $\pi^0$, the $\cos\theta$
distribution is flat in the range of [0,1]. Hence the $E_{low}$
energy spectrum is intrinsically flat, and the deviation from
flatness measures the photon inefficiency as a function of energy.
The distribution shows that the efficiency is low at low energy but
plateaus starting from 0.1 GeV.  A comparison of the photon energy
spectrum shape between data and Monte Carlo simulation shows that
the average difference is 0.6\%.  However, this measurement only
provides a relative efficiency as a function of photon energy.  The
absolute efficiency in the plateau region above 0.1 GeV can be
determined in data and simulation using high energy electrons from
radiative Bhabha events. The electromagnetic showers of electrons
and photons in the crystal calorimeter are the same for $E> 0.2$
GeV. The detection efficiency for electrons entering the EMC with
$E> 0.2$ GeV is essentially 100\% in both data and Monte Carlo
simulation, which indicates no significant systematic uncertainty in
the simulated efficiency of the EMC.
\par
The above three methods may be affected by photon conversions,
mainly in the Beryllium beam pipe and inner part of the MDC.  A
study using samples of $e^+e^-\to \gamma\gamma, \gamma \to e^+e^-$
events in which the converted photons are explicitly reconstructed
indicates systematic efficiency differences due to material in the
interaction region between data and simulation are negligible.
\par
Although each of the above three methods suffers from different
shortcomings, such as the resolution and tracking efficiency
dependence, they all give consistent results within 1\%, showing that
the photon efficiency uncertainty is less than 1\%.

\subsection{Tracking efficiency}
The tracking efficiencies for soft and hard pions are studied with
$\psi^\prime\to\pi^+\pi^- J/\psi, J/\psi\to l^+l^- (l=e,\mu)$ and
$J/\psi\to\rho\pi\to\pi^+\pi^-\pi^0$ event samples, respectively.  The
transverse momentum for a soft pion is less than 400 MeV$/c$. The
tracking efficiency is calculated with $\epsilon = N_{full}/N_{all}$,
where $N_{full}$ indicates the number of events of $\pi^+\pi^-l^+l^-
(\pi^+\pi^-\pi^0)$ with all final tracks reconstructed successfully;
$N_{all}$ indicates the number of events with one or both charged pion
tracks successfully reconstructed in addition to the lepton pair
($\pi^0$) for $\psi^\prime\to\pi^+\pi^- J/\psi, J/\psi\to l^+l^- $
($J/\psi\to\rho\pi\to\pi^+\pi^-\pi^0$). In addition, we require that
the direction of the missing momentum should be within the MDC
coverage. The missing momentum is calculated using the reconstructed
lepton-pair ($\pi^0$) and one of the reconstructed pions for the
$\psi^\prime\to\pi^+\pi^- J/\psi, J/\psi\to l^+l^- $
($J/\psi\to\rho\pi\to\pi^+\pi^-\pi^0$). A very clean soft pion sample
is selected by doing a kinematic fit with the lepton pair constrained
to the J/$\psi$ mass, and background from $J/\psi \rightarrow
\pi^+\pi^-$ is rejected using muon counter information and is
negligible. For the hard pion selected from
$J/\psi\to\pi^+\pi^-\pi^0$, the purity of the sample is more that
98\%; the small background is from $J/\psi\to K^*K\to K^+K^-\pi^0$ due
to misidentification of the kaon as a pion for the reconstructed
track. The differences for the soft and hard pion tracking
efficiencies between the data and MC are both estimated to be 2\%,
which is taken as the pion tracking uncertainty.

The kaon tracking efficiency is determined with a sample of $J/\psi
\rightarrow K^{*}(892)^{0}K^{0}_{S}+c.c\rightarrow
K^{0}_{S}K^{+}\pi^{-}+c.c\to K^+\pi^-\pi^+\pi^-+K^-\pi^+\pi^+\pi^-$
events.  The tracking efficiency is calculated in the same way as
the pion track efficiency. Here $ N_{full}$ is the number of events
with a matched kaon track in addition to the three tracks identified
as pions, and $N_{all}$ is the number of events with or without a
matched kaon and three tracks identified as pions. A clean sample is
selected by using a second vertex fit to $K_S$, and setting
stringent mass windows for the $K_S$ and $K^*(892)^0$. The
difference in the kaon tracking efficiency is about 2\% between the
data and MC, which is taken as the uncertainty of kaon tracking
efficiency.

\subsection{Efficiency for particle identification}
The efficiencies for pion and kaon PID are obtained with $J/\psi\to
\pi^+\pi^-\pi^0$ and $K^+K^-\pi^0$ control samples,
respectively. Samples with backgrounds less than 1\% are selected by
using a narrow $\pi^0$ mass window, and requiring one track be
identified as a pion (kaon) for $J/\psi \to \pi^+\pi^- \pi^0$
($K^+K^-\pi^0$) based on the TOF and dE/d$x$ information. The PID
efficiency is calculated with $\epsilon(PID) = N_0/(N^{\prime} +N_0)$,
where $N_0$ $(N^{\prime})$ denotes the events with the other track
identified (not identified) as a pion or kaon.  The differences
between data and MC for the pion and kaon PID efficiencies are about
2\%, and 2\% is taken as the systematic error.
\subsection{Uncertainties on the selection efficiencies}
The uncertainties on the selection efficiencies listed in
Table~\ref{tabsys} are considered as the systematic errors due to
the selection criteria. Control samples such as
$J/\psi\to\rho^0\pi^0$, $J/\psi\to\phi\eta$, and
$J/\psi\to\omega\eta$ are used to determine the efficiency
difference between data and MC for each selection criteria. For
$\psip\to\gamma\gamma\phi$, the uncertainty for the $\eta$ veto is
estimated to be 1.9\%, and that for the selection of the $\phi$
signal is 0.5\%. In $\psip\to\gamma\gamma\rho$, the dominant sources
are due to the electron and muon track rejection, the
$\cos\theta_{\gamma\gamma}$ requirement, the $\pi^0/\eta$ veto, and
the selection of the $\rho^0$ signal, which are $4.5\%$, $0.9\%$,
$1.2\%$, and $1.4\%$, respectively.  For this channel, the
uncertainty due to the muon track veto for both the $\pi^+$ and
$\pi^-$ (4.5\%) is determined from a study of a very pure $J/\psi
\to \rho^0 \pi^0$ sample.  In $\psip\to\gamma\gamma\omega$, the
uncertainty due to the background rejection and $\omega$ signal
selection is $1.4\%$.  The total uncertainties due to selection
criteria are 2.0\%, 5.0\%, and 1.4\% for $\psip\to\gamma\gamma\phi$,
$\psip\to\gamma\gamma\rho^0$, and $\psip\to\gamma\gamma\omega$,
respectively.
\par
Systematic errors in the fit to the $\gamma_{h} V$ mass distribution
originate from the uncertainties in the parameterizations for the
signal and background shapes. We obtain the uncertainties due to the
background shape by changing the order of the polynomial function
from 2 to 3 in the fit, and those for the signal shapes are
estimated by convolving a smearing Gaussian function to account for
the difference in the resolution between data and MC simulation.
Systematic errors associated with the bin size, fitting range, and
vector meson side-band regions are estimated by changing the bin
size, the fitting range, and the side-band regions.
\par
All errors are summarized in Table~\ref{tabsys}.  Finally, the total
systematic error varies from 8.8\% - 10.0\% depending on the final
state as summarized in Table~\ref{tabsys}.

\begin{table}
\caption{\label{tabsys} Sources of systematic errors ($\%$).}
\begin{ruledtabular}
\begin{tabular}{lccc}
 & $\gamma\gamma\phi$ & $\gamma\gamma\rho$ &$\gamma\gamma\omega$ \\
 & $\chi_{c0}$, $\chi_{c1}$, $\chi_{c2}$
       & $\chi_{c0}$, $\chi_{c1}$, $\chi_{c2}$
             & $\chi_{c0}$, $\chi_{c1}$, $\chi_{c2}$\\\hline
 Tracking & 4.0 &4.0 &4.0 \\
 PID& 2.0&2.0&2.0\\
 $\gamma$ detection &2.0&2.0&4.0\\
 4C-fit & 0.7& 0.7& --\\
 5C-fit &--&--& 3.1\\
 Uncert. of eff.& 2.0 &5.0&1.4\\
 BG shape &3.4, 2.6, 3.7 &2.4, 2.0, 2.8&3.2, 2.4, 4.1\\
 Binning &2.3, 1.5, 2.4&2.0, 1.0, 2.2&1.0, 0.0, 0.0\\
 Fit range and  &              &            &\\
 Sideband regions &2.1, 2.0, 2.5&1.0, 0.9, 1.5&1.0, 0.5, 1.0\\
 Signal shape &0.8&0.8&0.8\\
 No. of $\psip$ evts & 3.8&3.8&3.8\\
 ${\cal B}(\psip\to\gamma\chi_{cJ})$~\cite{pdg10}
      &3.3, 4.4, 4.0&3.3, 4.4, 4.0&3.3, 4.4, 4.0\\
 $V$ decay~\cite{pdg10}
      &1.0 &-- & 0.8 \\\hline
 Total &  8.8, 8.8, 9.3 & 9.5, 9.7, 10.0 & 9.3, 9.4, 9.8 \\
\end{tabular}
\end{ruledtabular}
\end{table}

\section{Helicity amplitude analysis }

In $\chi_{c1} \rightarrow \gamma V$ decays, the final state is a
superposition of longitudinal ($\lambda=0$) and transverse
($\lambda=\pm 1$) polarizations. The angular distribution is
$$\frac{d\Gamma}{\Gamma \,d \cos\theta} \propto (1-f_T)\cos^2\Theta +
 \frac{1}{2} f_T \sin^2\Theta,$$ where $f_T = |A_T|^2/(|A_T|^2 +
 |A_L|^2)$ is the transverse polarization fraction in the decay and
 $A_L$ and $A_T$ are the longitudinal and transverse polarization
 amplitudes, respectively, and $\Theta$ is the angle between the
 vector meson flight direction in the $\chi_{c1}$ rest frame and
 either the $\pi^+/K^+$ direction in the $\rho^0/\phi$ rest frame or
 the normal to the $\omega$ decay plane in the $\omega$ rest frame. By
 performing a likelihood fit to the angular distributions of the
 vector meson decays, we can determine the transverse polarization
 fraction $f_T$. We account for the different
 reconstruction efficiencies for the longitudinally and transversely
 polarized events using $f_T =
 f^{\rm{obs}}_T/(R+(1-R)f^{\rm{obs}}_T)$, where $f^{\rm{obs}}_T$ is
 the fraction of signal from transversely polarized signal events in
 data, and $R$ is the ratio of the longitudinal and transverse signal
 efficiencies.

\begin{figure}[htbp]
  \centering
   \includegraphics[width=0.35\textwidth,height=6.6cm,angle=0]{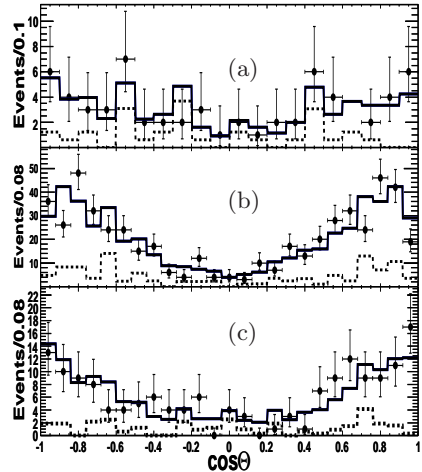}
             \put(-90,155){(a)}
             \put(-90,105){(b)}
             \put(-90,55){(c)}
\caption{\label{fig:heli}$\cos\Theta$ distributions and the fit for
(a) $\chi_{c1}\to \gamma\phi$, (b) $\chi_{c1}\to \gamma\rho^0$, and
(c) $\chi_{c1}\to \gamma\omega$. Dots with error bars are data; the
solid histogram is the fit; and dashed histogram is the sum of the
sideband background and the polynomial background.}
\end{figure}

In the likelihood fit, we take events in the $\chi_{c1}$ signal
region which is defined as $3.49\le M_{\gamma_h V}\le 3.52$~{\rm
GeV}/$c^2$. The signal shapes for the longitudinal and transverse
component are obtained from MC simulations. For the background
shapes, two sources are considered: one is the vector-meson sideband
background, which is normalized according to the number of sideband
events in the $\chi_{c1}$ signal region; the other background
considered is from MC simulated inclusive $\psip$ decay events (not
including signal events) that satisfy the selection criteria and
have an invariant mass in the $\chi_{c1}$ signal region, which are
normalized according the number of polynomial events used in the
$\chi_{c1}$ mass fit. The $\cos\Theta$ distributions are fitted with
the combined backgrounds and MC simulated transverse and
longitudinal signal shapes. The total signal yield and transverse
polarization fraction are floated in the fit. The fitted results are
shown in Fig.~\ref{fig:heli}.  The values of the fraction of the
transverse component are $0.29_{-0.12-0.09}^{+0.13+0.10}$ for
$\chi_{c1}\to\gamma\phi$, $0.158\pm0.034^{+0.015}_{-0.014}$ for
$\chi_{c1}\to\gamma\rho^0$, and
$0.247_{-0.087-0.026}^{+0.090+0.044}$ for
$\chi_{c1}\to\gamma\omega$, where the first errors are statistical
from the fit and the second ones are the systematic errors.

 Since $f_T$ is a ratio, many systematic errors cancel out, and only
 the effects due to binning of the $\cos\Theta$ distributions and the
 parameterization of the background shape (estimated by assuming the
 backgrounds (except for the vector-meson sideband
background) contribute entirely to either the longitudinal or the
 transverse component) are considered here. The uncertainties of the
 binning of $\cos\Theta$ are estimated to be $^{+15.4}_{-19.2}\%$,
 $^{+1.9}_{-0.6}\%$, and $^{+2.5}_{-0.5}\%$, and the parametrization of the
 background, which is the dominant
 systematic error, is estimated to be $^{+28.5}_{-23.1}\%$, $^{+9.3}_{-9.2}\%$
 , and $^{+17.6}_{-10.5}\%$ for
 $\chi_{c1}\to\gamma\phi$, $\chi_{c1}\to\gamma\rho^0$, and
 $\chi_{c1}\to\gamma\omega$, respectively.  In order to compare with
 CLEO-c's results, CLEO-c's $A_{\pm}/A_0$ have been used to determine
 $f_T$ as: $0.072_{-0.031-0.019}^{+0.041+0.002}$ for
 $\chi_{c1}\to\gamma\rho$, and $0.32_{-0.11-0.11}^{+0.17+0.05}$ for
 $\chi_{c1}\to\gamma\omega$. The results are consistent within
 $2\sigma$.

\section{Final Results and Discussion}

In summary, we present the measurements of radiative decays of
$\chicJ$ to light vector mesons. We find ${\cal B}(\chi_{c1}\to
\gamma\rho^0)=(228\pm 13\pm 16)\times 10^{-6}$ and ${\cal
B}(\chi_{c1}\to \gamma\omega)=(69.7\pm 7.2\pm 5.6)\times 10^{-6}$,
which agree with the results from the CLEO experiment~\cite{cleoc}.
We observe $\chi_{c1}\to \gamma\phi$ for the first time, and find
${\cal B}(\chi_{c1}\to \gamma\phi)=(25.8\pm 5.2\pm 2.0)\times
10^{-6}$. Upper limits at the 90\% confidence level on the branching
fractions for $\chi_{c0}$ and $\chict$ decays into these final
states are determined. The final results are listed in
Table~\ref{tabtotal}. The theoretical predictions for ${\cal
B}(\chi_{c1}\to\gamma V)$ including the hadronic loop contribution
in pQCD calculation~\cite{chen} are consistent with our measurements
within errors. In addition, the fraction of the transverse
polarization component of the vector meson in $\chi_{c1}\to \gamma
V$ decay is studied. Our measurements of the polarization of the
vector mesons indicate that the longitudinal component is dominant
in $\chi_{c1}\to\gamma V$ decay, as expected for an axial-vector
particle radiative decaying into a vector ($\phi$, $\rho^0$, and
$\omega$) in the framework of the vector dominance model taking into
account the Landau-Yang theorem~\cite{ves,yang1950}. This
observation may aid future development on the QCD calculation of the
partial waves in the $\chi_{c1} \rightarrow \gamma V$ decay.

\begin{table}
\caption{\label{tabtotal}Results on $\chi_{cJ}\to \gamma V$. The
upper limits are set at the 90\% C.L.}
\begin{ruledtabular}
\begin{tabular}{lccccc}
 Decay & No. of & Eff. & Syst.  & Br.   & Stat.\\
 mode  &  evts.   &(\%)& err.(\%) & ($10^{-6}$)   &sign.\\\hline
 $\chi_{c0}\to\gamma\phi$ &$15.0\pm6.6$ &32.4&8.8
                 &$<$16.2 &\\
 $\chi_{c1}\to\gamma\phi$  &$42.6\pm8.6$ &34.6&8.8
                 &$25.8\pm5.2\pm2.3$ &$6\sigma$\\
 $\chi_{c2}\to\gamma\phi$&$4.6\pm4.9$&32.6&9.3&$<$8.1&\\
 $\chi_{c0}\to\gamma\rho^0$ &$6 \pm12$ &22.6&8.1 & $<$10.5&\\
 $\chi_{c1}\to\gamma\rho^0$&$432\pm25$ &19.4&8.3& $228\pm13 \pm22$&$> 10\sigma$\\
 $\chi_{c2}\to\gamma\rho^0$&$13\pm11$ &15.7&8.7 & $<$20.8 &\\
 $\chi_{c0}\to\gamma\omega$&$5\pm11$&18.6&9.3&  $<$12.9 &\\
 $\chi_{c1}\to\gamma\omega$&$136\pm14$&22.7&9.4&$69.7\pm7.2\pm6.6$&$> 10\sigma$\\
 $\chi_{c2}\to\gamma\omega$&$1\pm6$&19.2&9.8&$<$6.1&\\
\end{tabular}
\end{ruledtabular}
\end{table}

\section{Acknowledgments}
The BESIII collaboration thanks the staff of BEPCII and the
computing center for their hard efforts. This work is supported in
part by the Ministry of Science and Technology of China under
Contract No. 2009CB825200; National Natural Science Foundation of
China (NSFC) under Contracts Nos. 10625524, 10821063, 10825524,
10835001, 10935007; the Chinese Academy of Sciences (CAS)
Large-Scale Scientific Facility Program; CAS under Contracts Nos.
KJCX2-YW-N29, KJCX2-YW-N45; 100 Talents Program of CAS; Istituto
Nazionale di Fisica Nucleare, Italy; Russian Foundation for Basic
Research under Contracts Nos. 08-02-92221, 08-02-92200-NSFC-a;
Siberian Branch of Russian Academy of Science, joint project No 32
with CAS; U. S. Department of Energy under Contracts Nos.
DE-FG02-04ER41291, DE-FG02-91ER40682, DE-FG02-94ER40823; University
of Groningen (RuG) and the Helmholtzzentrum fuer
Schwerionenforschung GmbH (GSI), Darmstadt; WCU Program of National
Research Foundation of Korea under Contract No. R32-2008-000-10155-0

\end{document}